\documentclass[12pt,preprint]{emulateapj}
\usepackage{natbib}
\begin{document}

\title{Dark energy and neutrino mass constraints from weak
lensing, supernova, and relative galaxy ages}

\author{Yan Gong$^{1,2}$, Tong-Jie Zhang$^{3,4}$, Tian Lan$^{3,1}$and Xue-Lei Chen$^{1,4}$}

\affil{ $^1$National Astronomical Observatories, Chinese Academy of
Sciences, Beijing 100012, China\\
$^2$Graduate School of Chinese Academy of Sciences, Beijing 100049, China\\
$^3$Department of Astronomy, Beijing Normal University, Beijing, 100875, China;tjzhang@bnu.edu.cn\\
$^4$Kavli Institute for Theoretical Physics China, CAS, Beijing 100190, China\\
}

\begin{abstract}
We use the current weak lensing data to constrain
the equation of state of dark energy $w$ and the total mass of
massive neutrinos $\sum m_{\nu}$. The constraint on $w$ would be weak if
only the current weak lensing data are used. With the addition of other observational
data such as the type Ia supernovae, baryon acoustic oscillation, and 
the high redshift Hubble parameter data $H(z)$ derived from relative galaxy ages
to break the degeneracy,
the result is significantly improved. For the pure $w$CDM model without massive neutrinos, 
we find $w=-1.00^{+0.10}_{-0.12}$. For the $w$CDM model with the massive
neutrino component, we show that the constraint on $w$ is almost
unchanged, there is very little degeneracy between $w$ and $\sum m_{\nu}$. 
After marginalizing over
other parameters, we obtain the probability distribution function of
$\sum m_{\nu}$, and find that the upper limit is $\sum m_{\nu} \leq 0.8$
eV at $95.5\%$ confidence level for the combined data sets. Our
constraints of $w$ and $\sum m_{\nu}$ are both compatible and comparable with the
constraints obtained from the WMAP 5-year data.

\end{abstract}


\maketitle

\section{Introduction}

Great progress is achieved in the study of modern cosmology in recent years. Thanks to
the precision observations of the ``distance indicator'' type Ia supernovae
(SN Ia) and the anisotropy of the cosmic microwave background (CMB), an
accelerating Universe with a non-baryonic dark matter component is 
established \citep{1998AJ....116.1009R,
1999ApJ...517..565P,2000Natur.404..955D,2003ApJS..148..175S}. Other
cosmological probes such as the large scale structures
\citep{2004PhRvD..69j3501T}, galaxy clusters \citep{2008MNRAS.383..879A},
Hubble rate derived from relative galaxy ages (RGA)\citep{2005PhRvD..71l3001S}, provide further
supporting evidence for this scenario, and help determine the model parameters with even
better precision.
However, the nature of the so-called dark energy which drives the
acceleration of the Universe is still far beyond our understanding. The simple
cosmological constant cold dark matter model ($\Lambda$CDM) remains to be the most 
popular, with an equation of state (EOS) $w\equiv p/\rho=-1$ for the dark energy. 
However, such a simple
cosmological constant suffers from the fine tuning and coincidence problems
\citep{1989RvMP...61....1W,1999PhRvL..82..896Z}. Dynamical dark
energy models typically have evolving EOS $w=w(z)$, a wide variety have been 
proposed and tested in literature \citep{2002PhLB..545...23C,
2005PhRvD..72l3515Z,2006PhLB..634..101F,2006PhRvD..73f3521X,
2007PhLB..648....8Z}. To eventually solve the dark energy problem,
it is very important to develop additional techniques and use more 
observational data to test the dark energy models and measure
other cosmological parameters.

Another fundamental problem in modern physics 
which we would like to draw particular attention to in this paper
is the mass of neutrinos. The neutrino oscillation experiments suggest that the mass
differences between various types of neutrinos are $\Delta
m_{12}^2\approx 8\times 10^{-5}$ eV$^2$ (from solar neutrino
experiment) and $\Delta m_{12}^2 \approx 2.2\times10^{-3}$ eV$^2$
(from atmospheric neutrino experiment) \citep{2004NJPh....6..122M}.
These results suggest that the masses of the neutrinos form a hierarchy of
$m_1\sim 0$, $m_2\sim\Delta m _{\rm solar}$ and $m_3\sim\Delta
m_{\rm atmospheric}$, or alternative an inverted order. However, it is also
possible for the three types of neutrinos have almost 
degenerate masses $m_1\sim m_2\sim m_3\gg \Delta m_{\rm
atmospheric}$. There are not absolute bounds on the neutrinos masses
from oscillation experiments, while the cosmological observations
can provide us effective ways to infer the abundance and total mass
of neutrinos, since the kinematics of neutrinos affects the
growth of cosmological structures \citep{2006ARNPS..56..137H}.
For the $\Lambda$CDM cosmology, upper bounds on the sum of the neutrino
masses have reached the level of $\sum m_{\nu}\lesssim 0.3-1$ eV at $2\sigma$,
depending on the data set used \citep[][and references
therein]{2006JCAP...06..025H}. If the dark energy EOS is taken into
account however, the upper bound is shown to be relaxed to about $1.5$ eV at
$2\sigma$ level \citep{2005PhRvL..95v1301H}. This may indicate a
degeneracy between the neutrino mass and dark energy EOS $w$
\citep{2005PhRvL..95v1301H, 2006PhR...429..307L}. The
inclusions of baryon acoustic oscillation (BAO) data
\citep{2006JCAP...06..019G} and the weak gravitational lensing
\citep{2006JCAP...06..025H} may help 
break such degeneracy, as we shall proceed to show in the remaining part of this paper.

The gravitational weak lensing (WL) effect generates small distortions
(of the order $1\%$) on the images of distant galaxies as the light passed through 
inhomogenous matter distribution along the line of sight. This effect can be 
measured statistically to yield information on 
the density field and the geometry of the Universe. A great advantage of 
WL is that it relies on the total matter content
of the Universe directly, thus the problem of galaxy-to-matter bias in the large
structure survey is avoided.
The WL surveys provide a powerful and precise probe of the dynamical
property of the Universe at redshift $z\lesssim 3$, and it is an
independent technique to measure the properties of dark energy
\citep{2008PhR...462...67M}. ``Cosmic shear'' is recently reported in galaxy WL
surveys, which provides important cosmological implications \citep{2000Natur.405..143W,
2000MNRAS.318..625B,2000A&A...358...30V,2000astro.ph..3338K}.

A direct measurement of the Hubble expansion rate at different redshifts 
can be obtained from the relative galaxy 
ages (RGA), i.e. the differential age of 
galaxies  which are passively evolving \citep{2002ApJ...573...37J}. 
This method was verified at low redshift with SDSS data \citep{2003ApJ...593..622J}, and
has been used in the reconstruction of scalar field potential of dark energy 
\citep{2005PhRvD..71l3001S} and the constraint on the cosmological model 
firstly\citep{2007MPLA...22...41Y}. 
It is potentially competitive with other methods which 
probe cosmic expansion history, such as SN Ia and BAO. Indeed, it measures $H(z)$ directly
while SN Ia measures only the distance and is related to $H(z)$ by integration, hence the
RGA method could be an even more sensitive probe in some cases\citep{2008arXiv0804.3135L}.

In this work, we focus on using WL data to 
constrain the equation of state of dark energy and the masses of neutrinos. With
the current WL survey data, the SN Ia data,  the
baryon acoustic oscillation (BAO) data, and the Hubble parameter $H(z)$ derived with the 
RGA method,  we constrain the EOS $w$ of the dark energy in a constant $w$ model, 
the sum of the neutrino mass $\sum
m_{\nu}$, and the other cosmological parameters. 
We also discuss the degeneracy of the parameters,
especially for $\sum m_{\nu}$ and $w$ with the different data sets. The CMB data are not included
in this work, as we wish to compare the results from the weak lensing
with that of the CMB data (WMAP 5-year measurement). We shall
assume the geometry of the Universe is flat in this work.

The outline of this paper is as follows. In Sec. 2 we present our methods and 
the observational data sets. The results are given in Sec. 3. Finally
we draw conclusions in Sec. 4.

\section{Methods and observational data}

In this section, we briefly introduce the theoretical predictions of
the observables. 
We employ the Markov Chain Monte Carlo (MCMC) technique to 
constrain the parameters. Eight MCMC chains
are generated for each combination of our data sets, and after the
convergence each chain contains about 100000 points 
to sample the probability distribution in the parameter space. Then, these
chains are thinned and joined together, and there are about 12000
points left to be utilized to perform the constraints
\citep{2008PhRvD..77j3511G}.

\subsection{Weak lensing}

As we know, the dark energy could affect the expansion rate and
the large scale structure (LSS) of the Universe, from observation of 
these two aspects of cosmological evolution we can measure
determine the property of dark energy very well. 
The WL provides us a powerful probe for both 
aspects \citep{2002PhRvD..65f3001H}. The
massive neutrinos could suppress the matter
power spectrum on small scales, due to their free streaming, thus 
reducing the convergence power spectrum of the weak lensing,
which is sensitive to the small scale matter distribution. 
Weak lensing is therefore a powerful measurement for both the dark energy and the
massive neutrinos.

For details of the WL technique the readers can refer to
\cite{2001PhR...340..291B}. We start with the power spectrum of the
convergence $\kappa$, which describes the strength of the lensing effect.
Under Limber's approximation, the convergence power spectrum $P_{\kappa}$
can be related to the matter power spectrum $P(k,z)$ as
\begin{eqnarray}
P_{\kappa}(l)&=&\frac{c}{H_0}\int_0^{z_s}\frac{W^2(z)}{r^2(z)E(z)}
P(l/r(z),z){\rm d}z \nonumber \\
&=&\frac{2\pi^2}{l^3}\int_0^{z_s}\frac{W^2(z)r(z)}{H_0E(z)}
\Delta^2(k,z){\rm d}z,
\label{pkappa}
\end{eqnarray}
where $\Delta^2(k,z)=\frac{k^3}{2\pi^2}P(k,z)$ is the dimensionless
matter power spectrum, $k=l/r(z)$ with $l$ the multipole and
$r(z)$ the comoving distance defined as $\frac{c}{H_0}\int_0^z
\frac{{\rm d}z}{E(z)}$, $H_0$ is the Hubble constant, and
$E(z)$ is the expansion rate of the Universe
\begin{equation}
E(z)=\frac{H(z)}{H_0}=\sqrt{\Omega_{m}(1+z)^3+(1-\Omega_{m})
(1+z)^{3(1+w)}},
\label{ez}
\end{equation}
where $\Omega_{m}$ is the matter density parameter and $w$ is the
EOS of dark energy. We give the detailed formula for the calculation of
$P_{\kappa}$ in the Appendix.
The shear correlation functions can be defined as
\citep{2008PhR...462...67M,2008PhRvD..77j3513D}
\begin{eqnarray}
\xi_+(\theta)&=&\frac{1}{2\pi}\int_0^{\infty}{\rm d}l\,lP_{\kappa}(l)J_0
(l\theta),\\
\xi_-(\theta)&=&\frac{1}{2\pi}\int_0^{\infty}{\rm d}l\,lP_{\kappa}(l)J_4
(l\theta),\\
\xi^{\prime}(\theta)&=&\xi_-(\theta)+4\int_{\theta}^{\infty}{\rm d}
\theta^{\prime}\frac{\xi_-(\theta^{\prime})}{\theta^{\prime}}-12\theta^2
\int_{\theta}^{\infty}{\rm d}\theta^{\prime}\frac{\xi_-(\theta^{\prime})}
{\theta^{\prime 3}},\\
\xi_E(\theta)&=&\frac{\xi_+(\theta)+\xi^{\prime}(\theta)}{2},\
\xi_B(\theta)\ =\ \frac{\xi_+(\theta)-\xi^{\prime}(\theta)}{2},
\end{eqnarray}
where $J_0$ and $J_4$ are the zeroth and forth order Bessel functions of
the first kind respectively.
The shear correlation functions can then be compared with the 
measurements directly. In this
work we use the $\xi_E(\theta)$ data from \citep{2007MNRAS.381..702B},
which contains two wide sky survey: the Canada-France-Hawaii Telescope
Legacy Survey of Wide fields \citep[CFHTLS-Wide, sky coverage 22
$\rm deg^2$,][]{2008A&A...479....9F} and the Red-Sequence Cluster Survey
\citep[RCS, sky coverage 53 $\rm deg^2$,][]{2002ApJ...577..595H}.

The WL observations play the key role in our constraint on the neutrino mass. 
We modify the code of calculating matter power spectrum to account for
the suppression due to neutrino mass, the details of our calculation 
can be found in the Appendix.

\subsection{Type Ia Supernovae}

The SN Ia are widely used as standard candles to measure the luminosity 
distance \citep{1998AJ....116.1009R,1999ApJ...517..565P}. The redshift-dependent
luminosity distances $d_L(z)$ of the SN Ia are determined by the expansion history and
geometry of the Universe.
In a spatially flat Friedmann-Robertson-Walker (FRW) Universe, the
luminosity distance with redshift $z$ is given by
\begin{equation}
d_L(z)=\frac{c(1+z)}{H_0}\int_0^z\frac{{\rm d}z'}{E(z')}. \label{dL}
\end{equation}
Then the distance modulus of the SN Ia can be written as
\begin{equation}
\mu_{th}(z)=5\log d_L(z)+25.
\end{equation}

We use here the SN Ia data recently published by the Supernova Cosmology
Project (SCP) team \citep{2008arXiv0804.4142K}. This data set
contains 307 SN Ia, selected from several current widely used
SN Ia data sets, including the  Hubble Space Telescope
\citep[HST,][]{2004ApJ...607..665R}, SuperNova Legacy Survey (SNLS)
\citep{2006A&A...447...31A} and the Equation of State: SupErNovae
trace Cosmic Expansion \citep[ESSENCE,][]{2007ApJ...666..694W}. They 
were re-analysed by the SCP team with the same procedure 
to get a consistent and high-quality ``Union'' SN Ia data set.

\subsection{Baryonic Acoustic Oscillation}

The acoustic oscillations in the 
plasma of the early Universe is imprinted on the matter
power spectrum.
This signatures in the large-scale clustering of galaxies
yield additional cosmological tests. Using a
large spectroscopic sample of 46748 luminous red galaxies
covering 3816 square degrees out to $z=0.47$ from the Sloan Digital
Sky Survey (SDSS), Eisenstein et al. (2005) successfully found the
acoustic peak in the matter power spectrum. The position of the feature
can be described by the model-independent $A$-parameter
\begin{equation}
A=\sqrt{\Omega_{m}}\left
[\frac{1}{z_1E^{1/2}(z_1)}\int_0^{z_1}\frac{{\rm d}z'}{E(z')}\right
]^{2/3},\label{A}
\end{equation}
with $z_1=0.35$ the redshift at which the measurement is taken
The value of the $A$ parameter is measured as $A=0.469\pm0.017$
\citep{2005ApJ...633..560E}.

If the neutrinos are taken into account, the SDSS constraint on BAO
can be approximated as \citep{2006JCAP...06..019G}
\begin{equation}
A=0.469\left(\frac{n_s}{0.98}\right)^{-0.35}(1+0.94f_{\nu})\pm0.017,
\label{bao}
\end{equation}
where $n_s$ is the primordial power spectrum of fluctuations (see the
Appendix), $f_{\nu}=\Omega_{\nu}/\Omega_{m}$ is the fraction of neutrinos
relative to matter density.

\subsection{$H(z)$ from relative galaxy ages}
The Hubble parameter $H(z)$ is related with the differential age of the
Universe by
\begin{equation}
H(z)=-\frac{1}{1+z}\frac{{\rm d}z}{{\rm d}t},\label{hz}
\end{equation}
so through the determination of ${{\rm d}z}/{{\rm d}t}$ it can be measured
directly. The later can be determined by using the differential
ages of passively evolving galaxies observed in the Gemini Deep
Deep Survey (GDDS). \cite{2005PhRvD..71l3001S} obtained a set of
9 $H(z)$ measurement in the redshift range $0\sim1.8$ using this method. 
Various cosmological models were tested using this data
set in the last few years
\citep{2007MPLA...22...41Y,2006ApJ...650L...5S,2007PhLB..644....7W,
2007Ap&SS.311..407Q,2008arXiv0804.3135L,2008JCAP...09..008C,
2008arXiv0807.0039F}. 
The constraint derived using this method is compatible and comparable to 
that of SN Ia \citep{2008arXiv0804.3135L,2008JCAP...09..008C,
2008arXiv0807.0039F}.

\section{Constraints on cosmological parameters}

In this section, we first study the case of pure $w$CDM model, which
contains a dark energy component with constant EOS $w$ and the cold dark
matter. The separate constraints for each data set are produced to
illustrate their power of constraints on the parameters and
the degeneracy directions. Next we include the massive neutrinos
component, first in the $\Lambda$CDM model (henceforth $\Lambda$NCDM model),  and then 
in the $w$CDM cosmology ($w$NCDM model). We give upper limits on $\sum m_{\nu}$ 
from the marginalized probability distribution function (PDF), 
and discuss the degeneracy between the $\sum m_{\nu}$ and $w$.

\subsection{EOS of dark energy in $w$CDM model}

\begin{figure}[!htb]
\centering
\includegraphics[scale = 0.4]{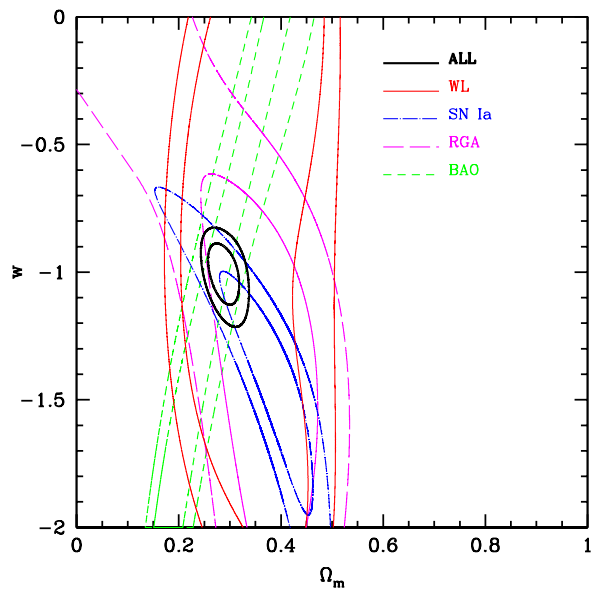}
\caption {The contour maps of $\Omega_{m}$-$w$ for
the WL (red solid), SN Ia (blue dash-dotted),
RGA (magenta long dashed), BAO
(green short dashed) and the combination constraint
(thick black solid). For each data set the $1\sigma$
(68.3\%) and $2\sigma$ (95.5\%) confidence levels (C.L.)
are shown respectively.}
\label{m0_w_W}
\end{figure}

The $1\sigma$ and $2\sigma$ contours of the $\Omega_{m}$-$w$ plane
are shown in Fig.\ref{m0_w_W}. We find that the WL data is  
sensitive to the matter density parameter $\Omega_{m}$, but less effective
in constraining the dark energy EOS $w$. Also,
the best-fit $\Omega_{m}$ is nearly independent of the $w$ for the WL data,
this is consistent with what were found in other works \citep{2006ApJ...647..116H,
2007MNRAS.381..702B}, so the degeneracy direction of the WL is
different from other cosmological observations. It is also
shown that the degeneracy directions of the SN Ia data and the
RGA data are almost the same, which again agrees with the conclusions 
reached in the recent works of 
\cite{2008arXiv0804.3135L,2008JCAP...09..008C,2008arXiv0807.0039F}.
The constraint from the SN Ia data is much tighter than that of the RGA data
due to the much larger sample of the former. If we combine all of the data sets, the
results are significantly improved. We measure from the global fit
of WL+SN Ia+RGA+BAO that, $\Omega_{m}=0.28^{+0.04}_{-0.03}$ and
$w=-1.00^{+0.10}_{-0.12}$, which are consistent with the recent
reported results from WMAP 5-year data \citep{2008arXiv0803.0547K}.
This result show that the cosmological constant is an excellent candidate of dark energy.

\subsection{Constraints on neutrino mass}

Now we investigate the constraints on neutrino mass for the
employed data sets. We use the WL, the WL+RGA+BAO, the WL+SN Ia+BAO and
the WL+SN Ia+RGA+BAO data set combinations to perform the constraints, so that
we could compare the power of parameter constraint
for the different data set combinations.

\begin{figure}{htb}
\begin{center}
\includegraphics[scale=0.4]{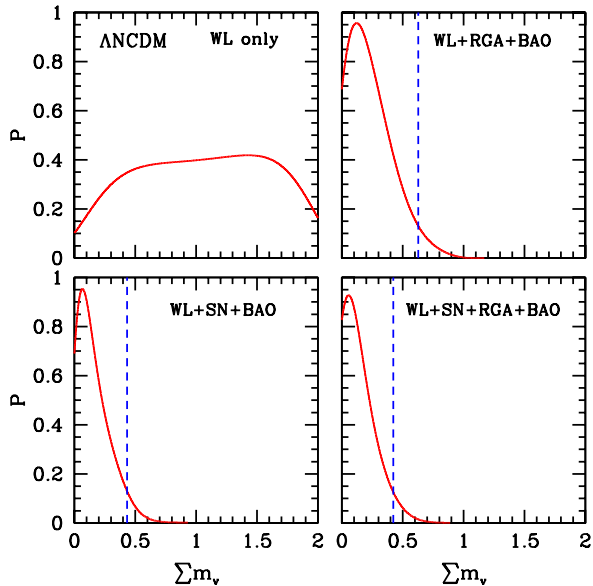}
\caption{The marginalized PDF of $\sum m_{\nu}$ in $\Lambda$NCDM model.
The $2\sigma (95.5\%)$ C.L. is also shown by the blue dashed line.}
\label{mv_L}
\end{center}
\end{figure}

Due to parameter degeneracy, with WL only there is very little constraining power. 
With the addition of BAO and RGA data, a $2\sigma$ limit of 0.6 eV can be obtained.
With SN Ia instead of RGA the constraint is better (0.4eV). If all (WL, BAO, SN Ia and RGA) 
data are combined to make the constraint, it is further slightly improved. It is also 
notable that the peak of the PDF distribution is not at 0 eV but at about 0.1eV for the 
last three data set combinations.

\begin{figure}[!htb]
\centering
\includegraphics[scale = 0.4]{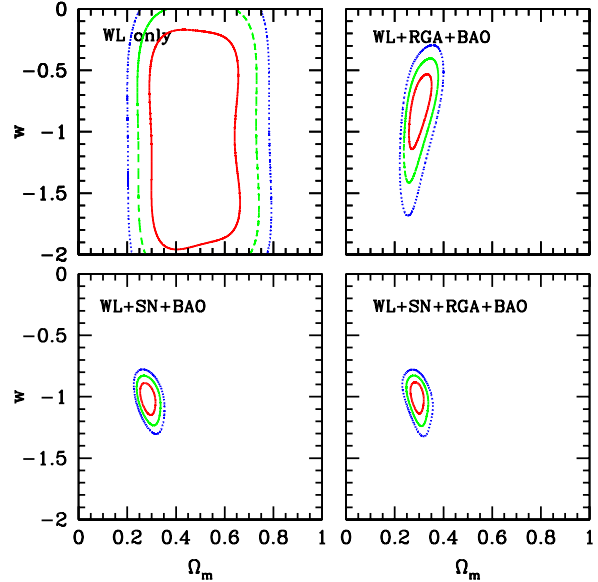}
\caption {The contour maps of $\Omega_{m}$-$w$ for
the different data sets.
The $1\sigma (68.3\%), 2\sigma (95.5\%)$
and $3\sigma (99.7\%)$ C.L. are marked by red solid,
green dashed and blue dotted lines respectively.}
\label{m0_w_M}
\end{figure}

In Fig. \ref{m0_w_M}, we show the contour map of the $\Omega_{m}$-$w$
plane in the $w$NCDM model. As can be seen, the results do not change much
compared with that of the $w$CDM model. The constraint of $w$ for
WL+RGA+BAO is looser than that for WL+SN Ia+BAO, but these two
combined data sets yield similar constraint on  $\Omega_{m}$.
This comparison actually reflects the constraining ability of the SN Ia 
data and the RGA data. As we shall also see in other
results presented below, except for $w$, the WL+RGA+BAO almost
have about the same constraining power as the WL+SN Ia+BAO 
data\citep{2008arXiv0804.3135L}, even though the number data points in the
RGA data set is much less than that of the SN Ia data set.

\begin{figure}[!htb]
\centering
\includegraphics[scale = 0.4]{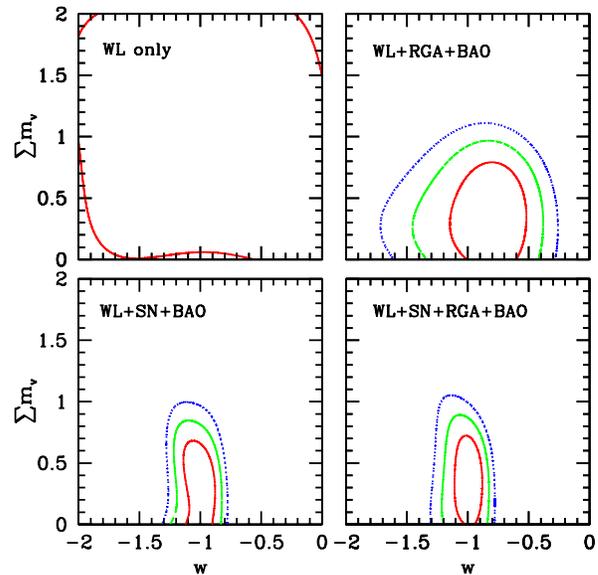}
\caption {The contour maps of $w$-$\sum m_{\nu}$ for
the different data sets.
The $1\sigma (68.3\%), 2\sigma (95.5\%)$
and $3\sigma (99.7\%)$ C.L. are marked by red solid,
green dashed and blue dotted lines respectively.}
\label{w_mv}
\end{figure}

The contour maps for $w$-$\sum m_{\nu}$ are shown in Fig. \ref{w_mv}.
The constraint is not good if only the WL data are employed.
After including the other data sets, the constraints of $\sum m_{\nu}$
are improved much, because the SN Ia, RGA and
BAO data could remarkably improve the constraints on the other
cosmological parameters such as $\Omega_{m}$ and $w$.
We can also see that $\sum m_{\nu}$ tends to be greater when the $w$
becomes more negative; however, there is no strong degeneracy between
$w$ and $\sum m_{\nu}$. The reason of this may be,  on one hand, the
current WL data are not yet accurate enough to indicate such relations; 
on the other hand, the $P_{\kappa}$, which is used to constrain
the parameters for the WL data, is the integral of the matter
power spectrum $P(k,z)$, so it is not as powerful as $P(k,z)$
to reflect the influence of $w$ on the formation of the LSS
\citep{2005PhRvL..95v1301H,2007ApJS..170..377S}.

\begin{figure}[!htb]
\centering
\includegraphics[scale = 0.4]{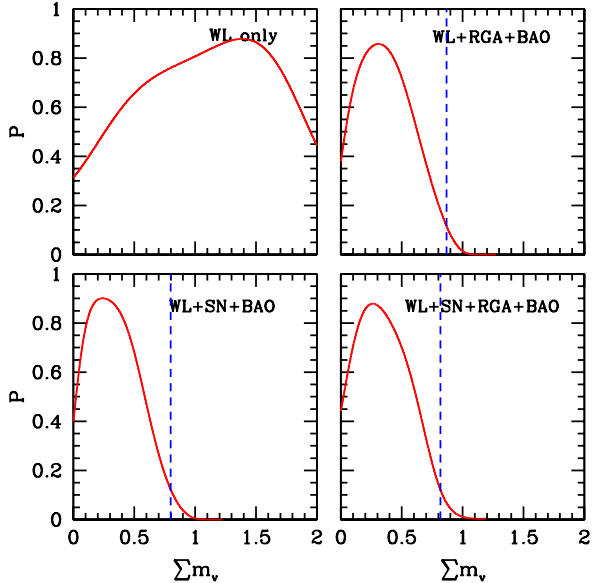}
\caption {The marginalized PDF of the $\sum m_{\nu}$
for the different data sets. The $2\sigma (95.5\%)$
C.L. is also shown by the blue dashed line.}
\label{mv}
\end{figure}

Finally, we marginalize over the other parameters and get the PDF of
$\sum m_{\nu}$ shown in Fig.\ref{mv}, the $2\sigma$ C.L. is also shown
as an vertical line.
We find $\sum m_{\nu}\leq 0.8$ eV at $95.5\%$ C.L. for the WL+SN Ia+BAO
and the WL+SN Ia+RGA+BAO data. This result is comparable with the recently
reported results $\sum m_{\nu}\leq 0.66$ eV from a combined analysis
of WMAP 5-year data together with SN Ia and BAO data
\citep{2008arXiv0803.0547K}. Moreover, as can be seen from this figure
the constraint range of $\sum m_{\nu}$ is fairly large if we just use the current
WL data, but the PDF for the WL+RGA+BAO data set is almost identical to
that for the WL+SN Ia+BAO, as we have discussed earlier.

\section{Conclusion and discussion}

In this work we use the WL survey data, combined with SN Ia,
RGA and BAO data 
to constrain the EOS of dark energy $w$ and the sum of
the neutrinos mass $\sum m_{\nu}$. The MCMC method is employed
to give the PDF of the parameters.

We first study the $w$CDM model without the neutrinos, and obtain the
constraint on $\Omega_{m}$-$w$ for different cosmological
observations. We find that the constraint for $w$ is weak if only the WL data
are used, but after inclusion of the other observational data, the
constraints are improved remarkably, and the cosmological constant
($w=-1$) is favored.

Then we considered constraints on the neutrino masses. In the $\Lambda$CDM model with 
massive neutrinos, WL along does not provide a constraint, but if one combine WL with
SN, BAO or RGA data, the neutrino mass is well constraint. The $2\sigma$ limit with 
all of these observations together is 
about 0.4 eV, and the peak of the fit is around 0.1 eV. We also considered the 
$w$CDM model with massive neutrinos. However, we find that $w=1$
(cosmological constant) is still favored in this model,
and a weak degeneracy is found between $w$ and $\sum m_{\nu}$. It
shows that WL can indeed break the degeneracy between $w$ and
$\sum m_{\nu}$ effectively \citep{2006JCAP...06..025H}.
After marginalizing over the other parameters we obtain the PDF for
$\sum m_{\nu}$. The upper limit of the neutrino mass
is obtained as $\sum m_{\nu} \leq 0.8$ eV at $2\sigma$, which is
comparable with the recent combined analysis of CMB+SN Ia+BAO data
\citep{2008arXiv0803.0547K}. Moreover, except for $w$, the constraint
ability for the WL+SN Ia+BAO and the WL+RGA+BAO
data set is nearly the same, which means that RGA data can play a similar
role in cosmological study as SN Ia \citep{2008arXiv0804.3135L,
2008JCAP...09..008C,2008arXiv0807.0039F}, the point of which is 
clearly demonstrated by \cite{2008arXiv0804.3135L}.

Although the capability of the current WL data on cosmological
parameter constraints is still not very good, and our constraints on
$w$ and $\sum m_{\nu}$ are not as strong as the other works
\citep{2005PhRvD..71d3511S,2005PhRvD..71j3515S,2006JCAP...10..014S,
2007PhRvD..75h3510K,2008arXiv0803.0547K,2008PhRvD..78c3010F}, it shows that
WL is a good and powerful complementary for the other observations,
and plays a more and more important role for the cosmological study.
When the next generation WL surveys, such as the SuperNova/Acceleration
Probe\footnote{{\tt http://snap.lbl.gov}} (SNAP) and the Large Synoptic Survey
Telescope\footnote{{\tt http://www.lsst.org}} (LSST), are put to work,
the WL observation would be more accurate and become an indispensable
measurement for the cosmology study.

\begin{acknowledgments}
Our MCMC chain computation was performed on the Supercomputing
Center of the Chinese Academy of Sciences and the Shanghai
Supercomputing Center. This work is supported by the Chinese
Academy of Sciences under grant KJCX3-SYW-N2, by the Ministry of 
Science and Technology National Basic Science program (project 973)
under grant No.2007CB815401 and No. 2009CB24901, by the Ministry of Education  
Science Research Foundation for the Returned Overseas Chinese Scholars, 
and by the National Science Foundation of China under the Distinguished Young Scholar
Grant 10525314, and Grants No.10473002, 10675019, 10503010.

\end{acknowledgments}

\appendix

\section{calculation of lensing power spectrum}

The weight function $W(z)$ in Eq.(\ref{pkappa}) is
written as \citep{2002PhRvD..65f3001H}
\begin{equation}
W(z)=\frac{3}{2}\left(\frac{H_0}{c}\right)^2\Omega_{m} f(z)(1+z),
\label{weight}
\end{equation}
with function
\begin{equation}
f(z)=r(z)\int_{z}^{\infty}\frac{r(z^{\prime})-r(z)}
{r(z^{\prime})}n(z^{\prime}){\rm d}z^{\prime},
\label{fz}
\end{equation}
in which $n(z)$ is the normalized number density distribution (i.e.,
$\int n(z){\rm d}z=1$) of source galaxies. The $n(z)$ we use here
is proposed in \cite{2007MNRAS.381..702B}, which takes the form as
\begin{equation}
n(z)=N\frac{z^a}{z^b+c},
\end{equation}
where $a$, $b$ and $c$ are free parameters, and $N$ is a normalizing
factor,
\begin{equation}
N=(\int^\infty_0 dz'\frac{z'^a}{z'^b+c})^{-1}.
\end{equation}


The linear matter power spectrum $\Delta^2_L(k,z)$ is parameterized as
\begin{equation}
\Delta^2_L(k,z)=Ak^{n_s+3}T^2(k)D^2(z),
\label{linear}
\end{equation}
where $D(z)=g(z)/(1+z)g(0)$ is the linear growth factor, $T(k)$ is the
transfer function, $A$ is the normalization factor and $n_s$ is the primordial
fluctuation spectrum. Hereafter we use a Harrison-Zel'dovich spectrum
$n_s=1$. For the $\Lambda$CDM model ($w=-1$), the relative growth factor
$g(z)$ is found to be well approximately by \citep{1992ARA&A..30..499C}
\begin{equation}
g_{\Lambda}(z)=\frac{(5/2)\Omega_{m}(z)}{\Omega_{m}^{7/4}(z)-\Omega_{\Lambda}
(z)+(1+\Omega_{m}(z)/2)(1+\Omega_{\Lambda}(z)/70)},
\end{equation}
with
\begin{equation}
\Omega_{m}(z)=\frac{\Omega_{m}(1+z)^3}{\Omega_{m}(1+z)^3+\Omega_{\Lambda}},\
\Omega_{\Lambda}(z)=\frac{\Omega_{\Lambda}}{\Omega_{m}(1+z)^3+\Omega_{\Lambda}}.
\end{equation}
For the transfer function, we adopt the fitting result of
\cite{1986ApJ...304...15B} for an adiabatic $\Lambda CDM$ model
\begin{equation}
T_{\Lambda}(q) = \frac{\ln(1+2.34q)}{2.34q}[1+3.89q+(16.1q)^2+
(5.46q)^3+(6.71q)^4]^{-1/4},
\end{equation}
where $q=k/h\Gamma$, and $h=H_0/(100 \,{\rm km}\,{\rm s}^{-1}\,
{\rm Mpc}^{-1})$,and $\Gamma=\Omega_{m}h\exp[-\Omega_{b}(1+\sqrt{2h}/
\Omega_{m})]$ is the shape parameter with baryon density $\Omega_{b}$.
$\Omega_{b}=0.0462$ is adopted according to the recent analysis of
CMB, SN Ia and BAO data \citep{2008arXiv0803.0547K}. For the extension
of the growth factor and transfer function from $\Lambda$CDM model
to any dark energy case with constant EOS $w$, we use the fitting form of
\cite{1999ApJ...521L...1M},
\begin{eqnarray}
g_Q &=& g_{\Lambda}(-w)^t,\\
t &=& -(0.255+0.305w+0.0027/w)[1-\Omega^w_m(z)] \nonumber\\
&&-(0.366+0.266w-0.07/w)\ {\rm ln}\ \Omega^w_m(z),
\end{eqnarray}
where $\Omega^w_m(z)=\Omega_{m}/[\Omega_{m}+(1-\Omega_{m})(1+z)^{3w}]$.
This fitting formula is accurate to $\rm 2\%$ for
$0.2\lesssim\Omega_{m}\le 1$ and $-1\le w\lesssim -0.2$.

For the non-linear power spectrum, we adopt the formula given by
\cite{1996MNRAS.280L..19P},
\begin{eqnarray}
\Delta_{NL}^2(k_{NL})&=&f_{NL}[\Delta_L^2(k_L)],\nonumber\\
k_L&=&[1+\Delta_{NL}^2(k_{NL})]^{-1/3}k_{NL},\nonumber\\
f_{NL}(x)&=&x\left[\frac{1+B\beta x+(Ax)^{\alpha\beta}}
{1+[(Ax)^{\alpha}g^3(z)/(Vx^{1/2})]^{\beta}}\right]^{1/\beta}.
\end{eqnarray}
The parameters in the non-linear function $f_{NL}$ are
\begin{eqnarray}
A&=&0.428(1+n_s/3)^{-0.947},\nonumber\\
B&=&0.226(1+n_s/3)^{-1.778},\nonumber\\
\alpha&=&3.310(1+n_s/3)^{-0.244},\nonumber\\
\beta&=&0.862(1+n_s/3)^{-0.287},\nonumber\\
V&=&11.55(1+n_s/3)^{-0.423},\nonumber
\end{eqnarray}
which are fitted to numerical simulation results.

\begin{figure}[!htb]
\centering
\includegraphics[scale = 0.35]{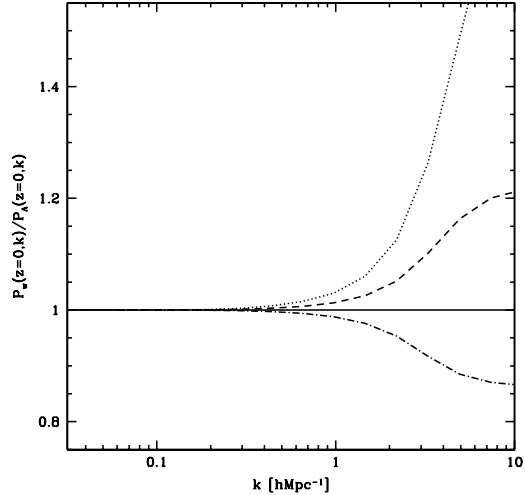}
\caption {The ratio of the $w$CDM matter power spectrum $P_w$
to the $\rm \Lambda CDM$ matter power spectrum $P_{\Lambda}$
as a function of $k$. The $P_w$ is calculated by our non-linear
power spectrum code. The dash-dotted, dashed and dotted lines
represent $w=-0.5, -0.75$ and $-1.5$ respectively,
and we set $\Omega_{m}=0.281$ which is to match the
parameter value for the black solid line in Fig. 1 of
\cite{2006MNRAS.366..547M}. We find our results match well
with that of \cite{2006MNRAS.366..547M} when $w<-0.5$.
Even for $w=-0.5$, our $P_w/P_{\Lambda}$ is about 1.6
at $k=10\, \rm h\, Mpc^{-1}$ while it is about 1.35 in
\cite{2006MNRAS.366..547M}, that still does not deviate much.}
\label{PQPL}
\end{figure}

Actually, the non-linear matter power spectrum we employ can
be seen as the Peacock-Dodds (PD96) fitting formula with
a modified growth factor proposed by \citep{1999ApJ...521L...1M}.
There is not yet a reliable analytical fit of the non-linear
power spectrum for the $w$CDM model \citep{2006ApJ...647..116H},
so now it is usually obtained from fitting to the N-body simulations,
the applicable range in the parameter space is always small.
For instance, an accurate fitting
non-linear matter power spectrum with varying $w$ was proposed
by \cite{2006MNRAS.366..547M}, but unfortunately, its applicable range
was $\Omega_{m} \in [0.211, 0.351]$. However, 
We have tested our power spectrum and find it matches well with
the simulation results of \cite{2006MNRAS.366..547M}, especially for
$w<-0.5$ (see Fig.\ref{PQPL}).

If there is a fraction of massive neutrinos in the matter components of
the Universe, the growth of the structure is suppressed by the
free streaming of neutrinos. The transition scale is 
the horizon scale when the neutrinos become
non-relativistic: $k_{nr}\approx 0.026\left(\frac{m_{\nu}}{1{\rm eV}}
\right)^{1/2}\Omega_{m}^{1/2}h\,{\rm Mpc}^{-1}$, below which
\citep{1998ApJ...498..497H}
\begin{equation}
\frac{\Delta P_L}{P_L}\approx -8\frac{\Omega_{\nu}}{\Omega_{m}},
\end{equation}
where $\Omega_{\nu}=\sum m_{\nu}/(93.2{\rm eV}\,h^2)$ is the neutrino
matter density \citep{2006JCAP...06..019G}. This approximation is shown to
be accurate for the linear theory regime with relative large scales
$k\lesssim 0.2h$ Mpc$^{-1}$ and small neutrino fraction $f_{\nu}$
\citep{2008JCAP...08..020B}. Hence, When we consider the
neutrino component the linear matter power spectrum should be modified
as $P_L^{\nu}=P_L+\Delta P_L \ (k>k_{nr})$. We add this modification
into our non-linear matter power spectrum code to constrain the sum
of the neutrino mass $\sum m_{\nu}$.

\bibliography{ztjcos-lensbib}
\bibliographystyle{aa}
\end{document}